\documentclass[preprint,floatfix,showpacs,aps]{revtex4}

\newcommand {\bea}{\begin{eqnarray}}
\newcommand {\eea}{\end{eqnarray}}
\newcommand {\be}{\begin{equation}}
\newcommand {\ee}{\end{equation}}

\usepackage{graphicx}
\usepackage{epsfig,subfigure}
\usepackage{amsmath}
\begin{document}
\def\({\left(}
\def\){\right)}
\def\[{\left[}
\def\]{\right]}

\def\Journal#1#2#3#4{{#1} {\bf #2}, #3 (#4)}
\def\RPP{{Rep. Prog. Phys}}
\def\PRC{{Phys. Rev. C}}
\def\PRD{{Phys. Rev. D}}
\def\FP{{Foundations of Physics}}
\def\ZPA{{Z. Phys. A}}
\def\NPA{{Nucl. Phys. A}}
\def\JPG{{J. Phys. G Nucl. Part}}
\def\PRL{{Phys. Rev. Lett}}
\def\PRpt{{Phys. Rep.}}
\def\PLB{{Phys. Lett. B}}
\def\AP{{Ann. Phys (N.Y.)}}
\def\EPJA{{Eur. Phys. J. A}}
\def\NP{{Nucl. Phys}}  
\def\ZP{{Z. Phys}}
\def\RMP{{Rev. Mod. Phys}}
\def\IJMPE{{Int. J. Mod. Phys. E}}

\title{Instabilities of relativistic mean field models and the role of nonlinear terms}

\author{A. Sulaksono$^{1}$, T. Mart$^{1}$,  T. J. B{\"u}rvenich$^{2}$, and J. A. Maruhn$^{3}$}

\affiliation{$^1$Departemen Fisika, FMIPA, Universitas Indonesia,
Depok, 16424, Indonesia\\
$^2$Frankfurt Institute for Advanced Studies, Universit\"at Frankfurt,
60438 Frankfurt am Main, Germany\\
$^3$Institut
f\"ur Theoretische Physik II,Universit\"at Frankfurt, 60438 Frankfurt am Main, Germany }

\begin{abstract}
 The  instability of nuclear matter due to particle-hole excitation
modes has been studied in the frame-work of several
relativistic mean field (RMF) models. It is found that both the
longitudinal and the transversal modes depend sensitively on
the parameter sets used. The important impact of the vector and
vector-scalar nonlinear terms on the stability of both modes
is demonstrated. Our finding corroborates the result of previous studies, namely that
certain RMF models cannot be used in high density
applications. However, we show that for certain parameter sets of RMF models  this shortcoming can be
alleviated by adding these
nonlinear terms.  
\end{abstract}
\maketitle

Remarkable progress has been made in constraining
the nuclear equation of state (EOS) from astrophysics and heavy ion
reactions, i.e., we know at a confident level that the EOS should
be soft at moderate densities but relatively stiff  at high densities
(for recent paper see, e.g., Ref~\cite{Klahn}). However, the
fast development in the unstable nuclear beam facility~\cite{Mueller} will also
reveal a lot of unexpected phenomena in unstable nuclei
far from the stability line region in the near future. Therefore, a unified model
describing simultaneously finite nuclei and matter properties at high
densities with high degree of accuracy is not only challenging but also
mandatory. 

Relativistic mean field (RMF) models have been quite successful in
providing a microscopic description of many  ground states properties
ranging from medium,
heavy, up to super-heavy nuclei (for a review see, 
e.g., Ref~\cite{PGR1}). The standard ansatz (S-RMF) uses $\sigma$, $\omega$ and $\rho$ mesons
as degrees of freedom with additional cubic and quartic
nonlinearities of $\sigma$ meson to describe the interaction. The
corresponding parameter
sets of this model are known as, e.g., NL-Z~\cite{Rufa} and
NL3~\cite{Lala}. The simplest extension of the S-RMF is
achieved by introducing a quartic nonlinearity of the $\omega$ meson in
the Lagrangian (V-RMF). The parameter sets TM1~\cite{Toki} and PK1~\cite{Meng} belong
to this model version. Other extensions (E-RMF) of the S-RMF parameterization are  known as G1 and
G2~\cite{Furnstahl96}. The corresponding model is derived from the
effective field theory which allows for possible scalar-vector coupling
terms up to fourth order to be present in the Lagrangian.  Other interesting properties of the RMF models come from the fact that the relativistic
nature of the models and the properties of
nuclear matter at saturation are fulfilled while in the extrapolation to higher
densities the appearance of acausal behavior (the
speed of sound exceeds the speed of light) can be alleviated~\cite{Jha}.

\begin{figure}
\epsfig{figure=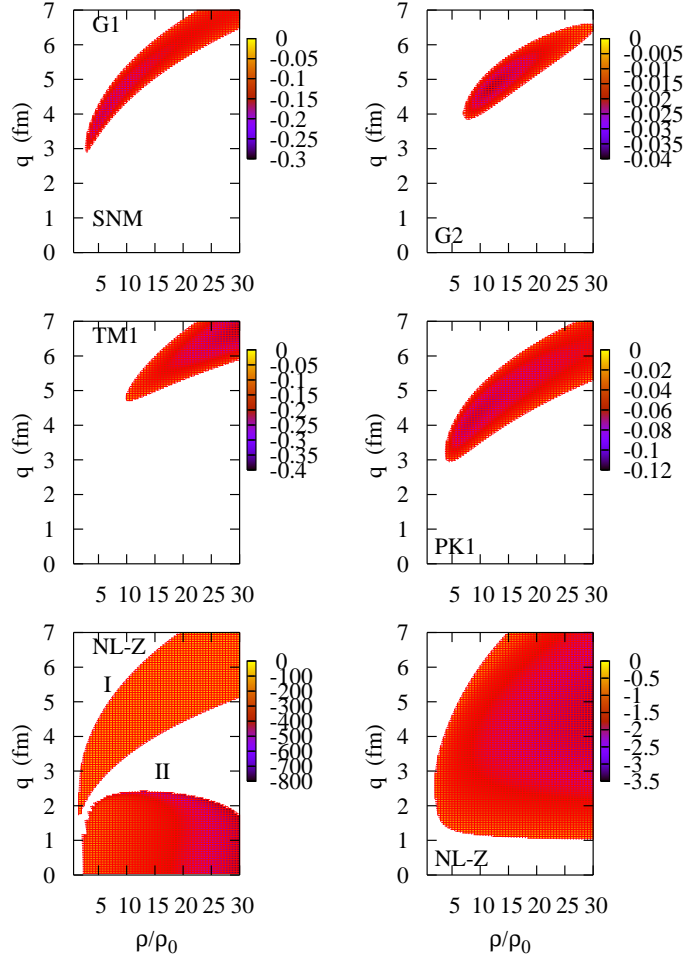,width=9cm}
\caption{(Color online) Longitudinal (left panels) and transversal
(right panels) modes of the instability due to the particle-hole
excitation with $q_0$ = 0 for the SNM and some selected RMF
parameterizations. Different depths of the dielectric functions are shown by
different colors (grey scales).}
\label{instab_LT}
\end{figure}

Surprisingly, though, little attention has been given so far to check the matter
instability by means of particle-hole excitations
with frequency $q_0$=0 in the transversal and longitudinal
modes, based on these models at high densities. It is understood that this analysis
is one of the tools to check the reliability of the models at high
densities. Efforts in this
direction have been devoted some times ago for the linear model case~\cite{Horo3,Fri},
whereas the latest progress can be found in Ref.~\cite{PGR2}, in
which the S-RMF model
was used to investigate the instability of symmetric
nuclear matter (SNM). The authors of Ref.~\cite{PGR2} found that the
onset of the instability depends sensitively on the
parameterization of the model. Parameterizations
with a low nucleon effective mass ($m^*/m$ $\sim$ 0.5-0.6), which can
accurately 
predict nuclear ground state properties, are critical in
the longitudinal mode. The S-RMF model produces also a wide
instability regime in the transversal mode, even for the parameterization
with high nucleon effective mass
or by using a more general functional form of nonlinear $\sigma$-meson self-interactions. In addition, it is known that at the critical density, which is
larger than saturation density, the $\sigma$ meson self-interaction in the S-RMF
model becomes unstable  because the square of the effective $\sigma$
meson mass becomes  negative and, as a consequence, an additional
instability regime in longitudinal
mode appears (see region II in the lower-left panel of Fig.~\ref{instab_LT}). This
instability is not present in the linear RMF model. To overcome the mentioned shortcomings, the
authors of Ref.~\cite{PGR2} suggested to add nonlinear terms which
contain 
not only functions of the $\sigma$ meson field, but also of the $\omega$ meson field. 

The analysis of the instability in symmetric nuclear matter (SNM)  is a first step
towards a more comprehensive analysis taking into account matter with multi-component
constituents, where the latter plays a key role in our understanding of
some stellar matter problems. Due to the simplicity of SNM, 
many essential physical aspects can be understood lucidly.  Therefore, we will revisit the instability problem of
the RMF models at high-density SNM but now
the role of vector and vector-scalar coupling nonlinearities are taken
into account and the effects in
the unstable regimes with respect to longitudinal and
transversal modes are investigated by using some selected
parameterizations of S-RMF, V-RMF and E-RMF models, where the  S-RMF
is used as the benchmark.  

To determine the instabilities of the RMF models, we start from the
energy density $\varepsilon$ in SNM which takes the following form,
\bea
\varepsilon&=&\varepsilon_{\rm linear}+\frac{1}{3} b_2\sigma^3+ \frac{1}{4} b_3\sigma^4\nonumber\\&-&\frac{1}{4} d_1  V_0^4 - d_2 \sigma V_0^2 - \frac{1}{2} d_3  \sigma^2 V_0^2,
\label{eq:edens}
\eea
where for the NL-Z and NL3 parameter sets (S-RMF) the last three terms vanish, and for
the TM1 and PK1 parameter sets (V-RMF) only the terms proportional to
$d_2$ and $d_3$ vanish, while for G1 and G2
parameter sets (E-RMF) all parameters are utilized. $\varepsilon_{\rm linear}$
is a function of the kinetic terms of the nucleons, $\sigma$ and $\omega$
masses as well as interactions terms of the nucleons. Using a similar procedure to the one
used in Refs.~\cite{Horo2,Horo3}, we can calculate the transversal
($\epsilon_T$) and
longitudinal ($\epsilon_L$) dielectric functions of the RMF models. For SNM,
they have simple forms, i.e.,
\bea
\epsilon_T&=&1+2 d_V^T \Pi_T\nonumber\\\epsilon_L&=&1+2 d_S
\Pi_S-2 d_V^L \Pi_V+4 d_{SV}^L \Pi_{SV},
\label{die}
\eea
with
\bea
\Pi_V &\equiv& \Pi_{00}-2 d_S\Pi_M^2+ 2d_S\Pi_S  \Pi_{00}\nonumber\\\Pi_{SV} &\equiv& \Pi_{M}+2 d_{SV}^L\Pi_M^2+ 2d_{SV}\Pi_S  \Pi_{00},
\eea
where $\Pi_T$, $\Pi_S$,  $\Pi_{00}$, and $\Pi_M$ are the transversal,
scalar, longitudinal and scalar-vector coupling polarizations 
 of proton
or neutron with $q_0=0$. In SNM, the values of each polarization
for proton and neutron are equal. The explicit forms of
these polarizations are given in Refs.~\cite{Horo2,Horo3}.
The longitudinal scalar meson propagator is given by
\bea
d_S=\frac{g_{\sigma}}
{q^2+m^{*~2}_{\sigma}+\Delta_{\sigma \omega}{(q^2+m^{*~2}_{\sigma})}^{-1}},
\eea 
while the vector meson longitudinal and transversal propagators are
\bea
d_V^L&=&\frac{g_{\omega}}{q^2+m^{*~2}_{\omega}+\Delta_{\sigma \omega}{(q^2+m^{*~2}_{\omega})}^{-1}}\nonumber\\d_V^T&=&\frac{g_{\omega}}{q^2+m^{*~2}_{\omega}},
\eea 
and the scalar-vector coupling propagator takes the form 
\bea
d_{SV}^L=\frac{g_{\omega}g_{\sigma} \Delta_{\sigma
\omega}}{(q^2+m^{*~2}_{\omega})(q^2+m^{*~2}_{\sigma})+\Delta_{\sigma
\omega}},
\eea
where $m_\sigma^{*~2}$= $\partial^2 \epsilon / {\partial\sigma}^2$,
$m_\omega^{*~2}$= $-\partial^2 \epsilon/{\partial V_0}^2$ and $\Delta_{\sigma \omega}$=$-\partial^2
\epsilon/\partial \sigma \partial V_0$. It is clear that the NL3, NL-Z,
TM1 and PK1 parameterizations have vanishing $\Delta_{\sigma \omega}$ and
only S-RMF parameterizations have a constant effective omega-meson
mass, which is equal
to its bare mass. The unstable regimes are determined from
$\epsilon_L$, $\epsilon_T$ $\le$ 0. The corresponding results for
some selected parameter sets are shown in Fig.~\ref{instab_LT}, where
the left
panels exhibit the longitudinal modes and the right panels display the
transversal ones.

The NL3 and NL-Z models have two instability regimes
(I and II). In regime II, NL3 has
a deeper $\epsilon_L$ valley compared to NL-Z and for each parameter set,
the  valley in regime II is deeper than regime I. Regime II of the NL-Z
parameter set starts to appear at $\rho_c$ $\sim$ 2.5  $\rho_0$, while
for NL3 it starts at $\rho_c$ $\sim$ 6.5  $\rho_0$, both with $q_c$
$\le$ 2.5 fm. If we use an artificial parameter set with a large nucleon effective
mass ($m^*/m$ $\ge$ 0.7 at saturation), the unstable regime II disappears. On the other hand,
for the V-RMF and E-RMF models with  $m^*/m$ $\sim$ 0.5-0.6 (the effective nucleon
mass range for which the correct spin-orbit splittings is reproduced) at saturation,
the regime II does not also exist. For NL-Z, regime I
appears quite early, i.e. at $\rho_c$ $\sim$ 2.5  $\rho_0$ with a
relatively small $q_c$ ($\sim$ 2 fm), while for NL3 it begins at  $\rho_c$ $\sim$ 5
$\rho_0$ with $q_c$ $\sim$ 3 fm. 

The vector and vector-scalar coupling in the nonlinear terms of the V-RMF and
E-RMF models lead to somewhat narrower unstable regimes compared to the standard one.
For TM1, it appears at $\rho_c$ $\sim$ 10
$\rho_0$ with $q_c$ $\sim$ 5 fm, for PK1 it starts at $\rho_c$ $\sim$ 5
$\rho_0$ with $q_c$ $\sim$ 3.5 fm, while for G2 it starts at $\rho_c$ $\sim$ 22
$\rho_0$ with $q_c$ $\sim$ 6.5 fm and for G1 it appears at $\rho_c$ $\sim$ 3
$\rho_0$ and $q_c$ $\sim$ 3 fm. This means that in the V-RMF and
E-RMF models there are parameter sets,
e.g. TM1 and G2, for which their  instabilities in the longitudinal mode can be shifted into the regimes
which are physically not too important (large $q_c$ and $\rho_c$)
without losing their accurate predictions in ground-state properties of
finite nuclei.  

\begin{figure}
\epsfig{figure=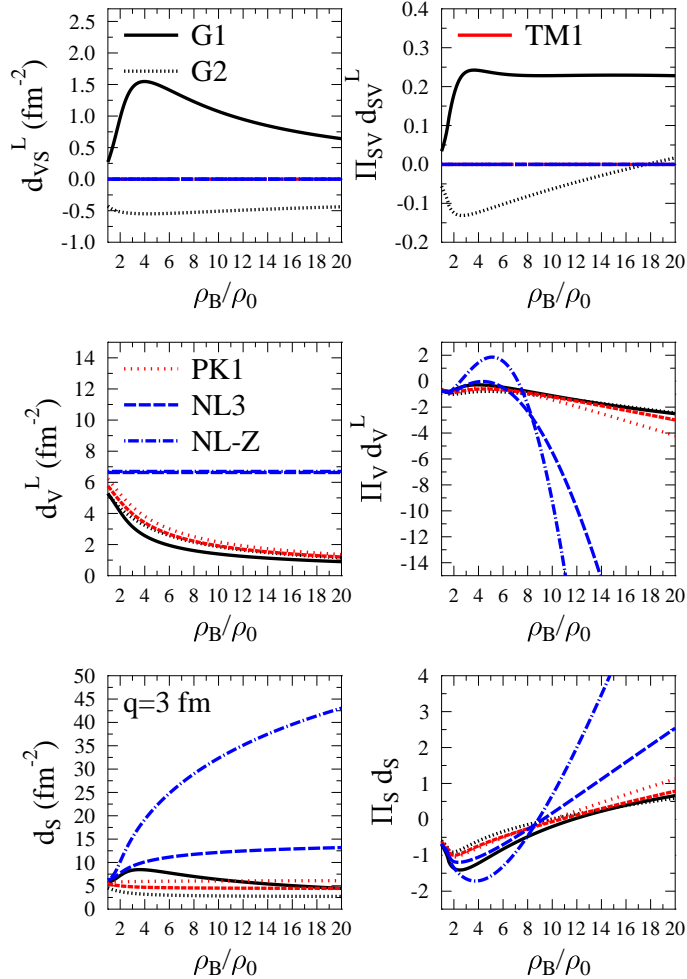, width=9cm}
\caption{(Color online) Longitudinal scalar ($d_S$), vector ($d_V^L$) and vector-scalar coupling
propagators ($d_{VS}^L$)(left panels) and the corresponding contributions to the longitudinal
dielectric function (right panels). Here $q$ = 3 fm
is used.}
\label{Apolar_1}
\end{figure}

Contributions of the longitudinal-scalar ($d_S$), vector ($d_V^L$) and
vector-scalar ($d_{VS}^L$)
propagators in the longitudinal dielectric function $\epsilon_L$ of
Eq.~(\ref{die}) with
$q$ = 3 fm (below  $q_c$ of 
 V-RMF and E-RMF models) for all
parameter sets used can be seen in Fig.~\ref{Apolar_1}. The density
dependence of the $\omega$ meson propagator in the V-RMF or E-RMF models
produces a negative vector contribution ($d_V^L \Pi_V$) for all
densities. In the case of the G1 parameter set, the  vector-scalar
coupling  ($d_{SV}^L \Pi_{SV}$) correction enhances the stability, but this
contribution is smaller than G1  albeit with different behavior for
G2. Overall, these yield a sufficient suppression to the negative contribution
from the scalar one ($d_S \Pi_S$) at low density regime. As a
consequence, a positive $\epsilon_L$ is produced. In
contrary, for the S-RMF model, a constant $\omega$ meson propagator allows  relatively
large positive vector and negative scalar contributions. This opens the possibility that
$\epsilon_L$ becomes negative at low densities.
\begin{figure}
\epsfig{figure=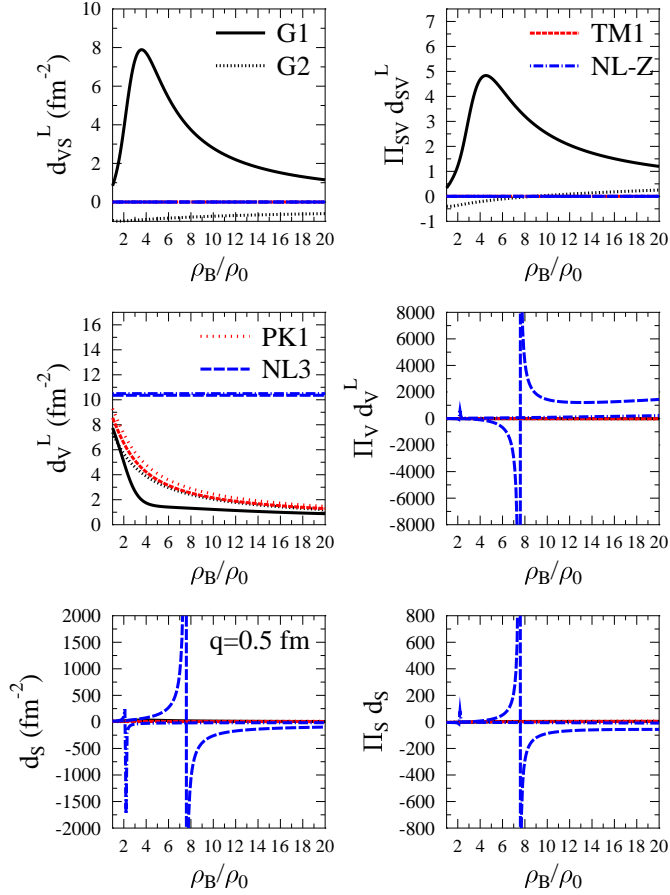, width=9cm}
\caption{(Color online) Same as Fig.~\ref{Apolar_1} but with $q$ = 0.5 fm. Different
with other parameter sets which have small regular contributions,
contribution of the NL3 and NL-Z parameter sets at certain
density scalar propagator exhibit discontinuities (lower left) which
leads to the appearance of the large
positive $d_V^L \Pi_V$ (middle right) and negative $d_S \Pi_S$
(lower right) above this density.} 
\label{Bpolar_2}
\end{figure}

\begin{figure*}[!]
\epsfig{figure=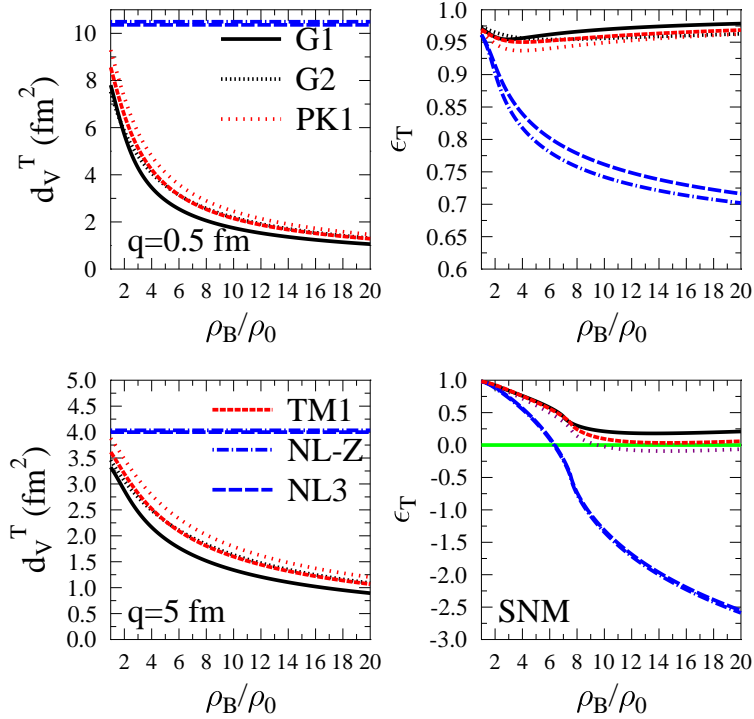, width=10cm}
\caption{(Color online) Transversal propagators with  $q$ = 5 fm (lower-left panel) and
$q$ = 0.5 fm (upper-left panel) followed by their transversal dielectric
functions (right panels).}
\label{transvers}
\end{figure*}
The same contributions but with  $q$ = 0.5 fm  can be seen in Fig.~\ref{Bpolar_2}. The square
effective $\sigma$ meson mass of the NL-Z and NL3 parameter sets becomes
negative after reaching a certain critical density ($\rho_c$) due to
particular parameter values of their scalar nonlinear terms, which leads to the appearance
of a discontinuity in $\sigma$ meson propagator at that point. For NL3, the effect is more dramatic 
but the regime is shifted to a higher critical density  
compared to the NL-Z parameter set. Furthermore, large positive vector and negative scalar
contributions for $\rho$ $\ge$ $\rho_c$ appear in this model. Therefore, in this density
range, the longitudinal mode becomes unstable, even for very small
momentum response ($q \sim$ 0). This is the reason that for both parameter sets the
unstable regime II exists.  
 
The S-RMF model has a broad
instability for this mode. It does not depend on the parameterization
used, except for low momentum response ($q$ $<$ 1 fm). The interesting finding
here is that for
TM1 and G1 parameter sets the unstable regimes absolutely
disappear. The particular parameter values of the vector and
vector-scalar nonlinear terms are responsible for this case.  While for the PK1 and G2
parameter sets relatively narrow unstable regime are produced, the regimes are also
shifted to
a relatively high density  and large momentum ( $\rho_c$ $\sim$ 5
$\rho_0$ with $q_c$ $\sim$ 3 fm for PK1 and  $\rho_c$ $\sim$ 8
$\rho_0$ with $q_c$ $\sim$ 4 fm for G2). This behavior never occurs for 
the S-RMF model parameterizations. The corresponding results for some selected parameter sets
are shown in the right panels of Fig.~\ref{instab_LT}. The role of the
density-dependent 
effective-transversal omega meson propagator of the V-RMF and E-RMF models
in stabilizing the transversal modes of the SNM for $q$ $\sim$ 0.5 fm and  $q$
$\sim$ 5 fm can be seen in Fig.~\ref{transvers}.

\begin{figure*}
\epsfig{figure=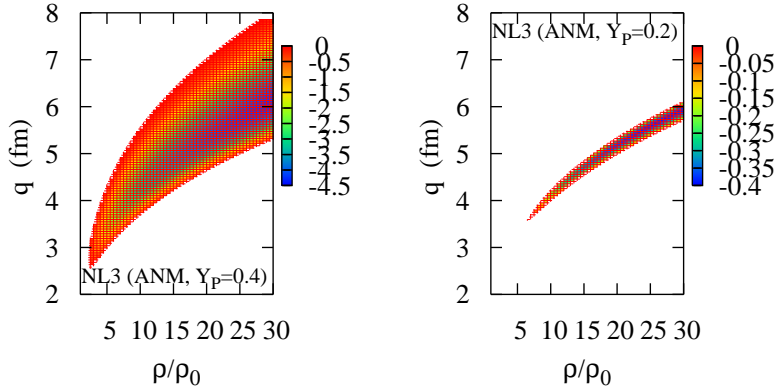, width=10cm}
\caption{(Color online) Unstable longitudinal regimes I for asymmetric nuclear matter
using NL3 parameter set with proton fraction equals 0.4 in the left panel and
0.22 in the right panel.}
\label{ANM}
\end{figure*}

If we extend this analysis to the asymmetric nuclear matter (ASM) case
(Fig.~\ref{ANM}),
it is found
 that
different with the unstable regime of transversal mode, the instability of
longitudinal mode depends sensitively on the proton fraction  in nuclear matter and after a certain
critical proton fraction, which is less than the neutron fraction, this
unstable regime disappears. 

Thus the longitudinal mode has a strong correlation with the
isovector sector of the model. In the case of multi-component matter
($p$, $n$, $e$, and $\mu$) in $\beta$ stability condition, a similar
instability trend of the longitudinal and transversal modes
with the ones obtained in the SNM case has been met. This means
that the nucleons
(protons and neutrons) play the main role for determination of the
instability. However, a fine tuning in the isovector
sector~\cite{Ant} can shift the instability of longitudinal mode
to higher critical density and momentum.   Details of these results will be reported elsewhere.
 
In conclusion, we have studied nuclear matter instability caused by particle-hole
excitations at $q_0$ = 0 in the RMF models at high density for
longitudinal and transversal modes. It is found that in both modes
the unstable regimes are very sensitive to the parameter set used. {\em This
opens the possibility to use the instability analysis as a tool to
explore the applicability of parameter sets of the existing RMF
models in high density applications}. For example, in transversal
modes, without the presence of vector and/or vector-scalar nonlinear
terms with particular parameter values, the unstable regimes can not vanish.  In general additional nonlinear terms in the form of vector and vector-scalar
coupling terms of the V-RMF and E-RMF models improve the stability of
both models at high densities. Finally we have observed that the longitudinal mode is  
sensitive  to
the isovector sector of the model. Thus, some important observables for neutron stars, like proton fraction and
asymmetry energy, can be related to this mode.
 
A.S. would extend his thanks to P-G. Reinhard for useful
discussions. A.S. and T.M. acknowledge the support from the Hibah Pascasarjana grant
as well as from the Faculty of Mathematics and Sciences, University of
Indonesia.

\end{document}